\documentclass[journal]{vgtc}                     

\onlineid{1159}

\vgtccategory{Research}

\vgtcpapertype{evaluation}

\title{Why Change My Design: Explaining Poorly Constructed Visualization Designs with Explorable Explanations}

\author{%
  \authororcid{Leo Yu-Ho Lo}{0000-0002-3660-3765},
  \authororcid{Yifan Cao}{0000-0002-5892-5052},
  \authororcid{Leni Yang}{0000-0003-4527-4905}, and
  \authororcid{Huamin Qu}{0000-0002-3344-9694}
}

\authorfooter{

  \item
    Leo Yu-Ho Lo, Yifan Cao, Leni Yang, and Huamin Qu are with the Hong Kong University of Science and Technology, Hong Kong.
    E-mail: \{yhload@cse, ycaoaw@connect, lyangbb@cse, huamin@cse\}.ust.hk.

}

\usepackage{graphicx}
\graphicspath{{figs/}{figures/}{pictures/}{images/}{./}} 

\usepackage{tabu}                      
\usepackage{booktabs}                  
\usepackage{lipsum}                    
\usepackage{mwe}                       

\usepackage{mathptmx}                  
\usepackage{multirow}

\newcommand{\ie}{\textit{i.e.}}
\newcommand{\etal}{\textit{et al.}~}
\newcommand{\etalcomma}{\textit{et al.},~}
\newcommand{\shorttext}{Short Text}
\newcommand{\longtext}{Long Text}
\newcommand{\correction}{Correction}
\newcommand{\correctionp}{Corrections}
\newcommand{\redraw}{Redraw}
\newcommand{\redrawp}{Redraws}
\newcommand{\highlighting}{Highlighting}

\newcommand{\annotation}{Annotation}
\newcommand{\annotationp}{Annotations}
\newcommand{\expexp}{Explorable Explanations}
\newcommand{\osf}{\href{https://osf.io/35spf}{https://osf.io/35spf}}
\newcommand{\squeezeafterfigure}{\vspace{-18pt}}
\newcommand{\squeezebetweenfigureandcaption}{\vspace{-20pt}}
\usepackage[normalem]{ulem} 
\ifx\revision\undefined
    \newcommand{\leo}[2]{{#2}}
    \newcommand{\leocw}[2]{{#2}}
    
    \newcommand{\leonotice}[1]{\empty}
\else
    \newcommand{\leo}[2]{{\sout{#1}\color{blue}{#2}}}
    \newcommand{\leocw}[2]{{\sout{#1}\color{orange}{#2}}}
    
    \newcommand{\leonotice}[1]{\leo{}{#1}}
\fi

\abstract{%
  Although visualization tools are widely available and accessible, not everyone knows the best practices and guidelines for creating accurate and honest visual representations of data. Numerous books and articles have been written to expose the misleading potential of poorly constructed charts and teach people how to avoid being deceived by them \leocw{and}{or} making their own mistakes. These \leocw{written materials}{readings} use various rhetorical devices to explain \leocw{these}{the} concepts to their readers. In our analysis of a collection of book\leocw{}{s}, online materials, and a design workshop, we identified six common explanation methods. To assess the effectiveness of these methods, we conducted two crowdsourced studies (each with $N=125$) to evaluate their ability to teach and persuade people to make design changes. In addition to these existing methods, we brought in the idea of \leocw{explorable explanations}{\expexp}, which allows readers to experiment with different \leocw{}{chart} settings and observe \leocw{the effects on visualizations.}{how the changes are reflected in the visualization.}
  While we did not find significant differences across explanation methods, the results of our experiments \leo{demonstrate that participants were better able to identify misleading charts after learning the explanations and accepting suggestions with explanations.}{indicate that, following the \leocw{acquisition of}{exposure to} the explanations, the participants showed improved proficiency in identifying deceptive \leocw{graphs}{charts} and were more receptive to proposed alterations of the visualization design.}
  \leocw{In the persuasiveness assessment, we discovered that participants were \leocw{inclined}{willing} to accept more than 60\% of the proposed adjustments.}{We discovered that participants were willing to accept more than 60\% of the proposed adjustments in the persuasiveness assessment.} Nevertheless, we found no significant differences among \leocw{the}{different} explanation methods in convincing participants to accept the modifications.
}


\begin{document}


\firstsection{Introduction}

\maketitle


The visualization community has developed guidelines for practitioners, outlining best practices and common pitfalls to avoid. These guidelines, such as starting the Y-axis at zero when drawing a bar chart and selecting sequential color schemes for continuous variables, are frequently emphasized. In addition, numerous books written by authors such as Huff \cite{huff1993lie}, Tufte \cite{tufte1983visual}, Cairo \cite{cairo2019charts}, \leocw{}{and} Monmonier \cite{monmonier2018lie}, as well as many other articles \cite{szafir2018good,correll2017black,pandey2015deceptive}, have addressed how poorly constructed visualizations can mislead the audience.

The major \leocw{takeaways}{takeaway} \leocw{in}{from} these readings \leocw{focus on equipping readers}{is that readers are equipped} with the skills necessary to (1) identify misleading visualizations\leocw{}{,} and (2) avoid creating deceptive visualizations themselves. These two abilities encompass the dual roles of being both consumers and producers of data visualizations in everyday life. A common rhetorical strategy in these works involves presenting an example of a misleading visualization and then systematically explaining why it is deceptive. This emphasis on explanation is crucial in educating readers \leocw{on recognizing}{to recognize} misleading visualizations \leocw{produced by others}{others produced} and \leocw{avoiding}{avoid} making similar mistakes in their own work.

Explaining \leocw{}{data} visualization best practices is also crucial to automatic visualization recommendation and correction systems. Early efforts in detecting visualization errors primarily focused on finding the issues within a given visualization, typically supplying an error message as the output \cite{chen2021vizlinter,mcnutt2018linting,mcnutt2020surfacing}. While it is important for these automated systems to detect problems and potentially suggest solutions, the communication aspect is often overlooked. 

To effectively integrate these diagnostic systems into existing visualization tools and assist creators, it is necessary to address the communication gap. Presenting users with clear explanations of why their visualization design may be potentially misleading and persuading them to accept the suggested fix is a challenge that must be tackled before these systems can be widely adopted and trusted by \leocw{}{the} end users.

In this work, we investigate practical methods for delivering explanations to readers. We begin by reviewing the literature and examining examples collected from the \leocw{Internet}{internet} to summarize how people explain misleading visualizations to their audiences. Subsequently, we organized an explanation formulation workshop to gather ideas on elucidating deceptive visualizations and to apply the explanation techniques derived from our literature review and online sources.

Through this process, we \leo{identify}{identified} six techniques for explaining misleading visualizations, which can serve as a foundation for improving the communication of \leocw{}{data} visualization best practices and enhancing the effectiveness of automatic visualization recommendation and correction systems.

In addition to the existing explanation techniques, we propose adopting Bret Victor's concept of \leocw{explorable explanations}{\expexp} \cite{victor2011explorable} as an interactive method for conveying information to readers. \leocw{Explorable explanations}{\expexp} enable users to engage with and investigate complex concepts, ideas, and data in an interactive and immersive manner. This approach is well-suited to address the explanatory challenges faced by readers and, potentially, visualization tool users\leocw{, providing}{. It also provides} an engaging and interactive \leo{way to understand and rectify misleading visualizations}{medium to comprehend and reassess the design choices of potentially deceptive visual representations}.

We conducted two between-subject experiments on crowdsourcing platforms to evaluate the learnability and persuasiveness of five explanation techniques for addressing various visualization mistakes. In the learnability experiment, we measured participants' prior\leocw{}{-} and post\leocw{ }{-}intervention performance in identifying misleading visualizations to assess the effectiveness of the methods.
For the persuasiveness experiment, we evaluated the acceptance rate of the suggested visualization corrections to determine the impact of the explanation techniques on users' willingness to adopt the proposed changes.


Through this work, we aim to contribute the following to the visualization community: (1) a comprehensive compilation of existing explanation techniques for addressing misleading visualizations\leocw{,}{;} (2) a prime example of \leocw{explorable explanations}{\expexp} focused on clarifying deceptive visualizations\leocw{,}{;} and (3) two evaluation experiments assessing the learnability of identifying misleading visualizations and the persuasiveness of accepting suggested corrections.

\section{Related Work}

Our study builds on the \leo{pioneer researches}{pioneering research} on misleading visualizations, also known as deceptive visualizations. Previous studies collected the existence of misleading visualizations and coined the term ``lies'' to describe them. Recent work focuses on developing algorithms to \leocw{automatic}{automatically} detect\leo{ the}{} them when the users \leocw{making}{are creating} visualizations using \leocw{}{computer} programming tools. \leo{Limited work on how to present and explain the issues to the users or readers.}{\leocw{However}{Yet}, there is a significant gap in \leocw{}{the} research when it comes to assessing the effectiveness of different explanation methods in making these issues clear to users or readers.}

\subsection{Misleading Visualizations}

The two most influential books on this topic are Huff's \textit{How to Lie with Statistics} \leo{}{in 1954} \cite{huff1993lie} and Tufte's \textit{The Visual Display of Quantitative Information} \leo{}{in 1983} \cite{tufte1983visual}. Both books \leocw{address}{addressed} the issue of misleading charts, using examples gathered from \leocw{}{the} news media of their \leocw{time}{respective times}. The deceptive tactics discussed in these books remain prevalent and widely debated. Tufte's introduction of the Lie Factor, a formula for identifying misleading charts, is a unique example of a quantitative approach to the problem.\leo{ Comparable instances include the 45-degree banking rule for line charts \mbox{\cite{heer2006multi}} and the aspect ratio of rectangular treemaps \mbox{\cite{kong2010perceptual}}, although these rules are rarely applied in practice.}{}
Other explanation techniques featured in these classic books, such as correction and annotation, continue to be widely employed when discussing visualization pitfalls and best practices.

Monmonier authored \textit{How to Lie with Maps} \cite{monmonier2018lie} from a cartographer's perspective, \leocw{which specifically addresses misleading presentation}{specifically addressing the misleading presentation} of information and geographic data in maps. \leocw{Jones'}{Jones's} book, \textit{How to Lie with Charts} \cite{jones1995lie}, offers a creator's perspective on avoiding the production of deceptive charts when using spreadsheet\leocw{}{s} and slideshow software. Cairo's \textit{How Charts Lie} \cite{cairo2019charts} provides a modern take on the topic, featuring recent examples of misleading techniques collected from social media and news media.\leo{ Bergstrom and West's book, \mbox{\textit{Calling Bullshit} \cite{bergstrom2021calling}}, and their online course bearing the same attention-grabbing title, adopt a data skepticism viewpoint in the big data era.}{}\leonotice{[Moved to a later paragraph of this subsection.]}
While these works expand upon the issues initially discussed by Huff and Tufte by examining a broader range of charts beyond printed media, the techniques used in their explanations remain similar.

Misinformation is widespread on the \leocw{Internet}{internet}, with data visualization often serving as a key medium for disseminating such inaccurate information. Several studies have investigated the use of visualizations in the context of misinformation online. Lo \etal aimed to broaden the scope of poorly constructed visualizations, addressing \leocw{}{a wide variety of} deceptive techniques that result in misleading representations and mistakes that render visualizations uninformative \cite{lo2022misinformed}. Lee \etal analyzed social media to understand the role of visualizations in social media posts and the dissemination of misinformation \cite{lee2021viral}. Lisnic \etal collected and examined instances of visualizations being used to spread misinformation, irrespective of the \leocw{quality}{deceptiveness} of the \leocw{chart's construction}{charts} \cite{lisnic2023misleading}.

The deceptive impact of misleading visualizations has been consistently demonstrated across various studies. Pandey \etal evaluated the effects of message exaggeration/understatement and message reversal through a crowdsourced experiment\leocw{, and their}{. Their} results revealed that participants who viewed the misleading \leocw{chart versions}{version of the chart} interpreted the data differently \cite{pandey2015deceptive}. In another study, Correll \etal focused on different visual designs of truncated axes trying to mitigate the misleading effect of \leocw{}{a} truncated axis \cite{correll2020truncating}. Their findings indicated that, despite attempts to mitigate the issue by hinting at the presence of a truncated axis, truncation consistently led to an exaggerated perception of the differences between bars.

\leo{}{Critical thinking is an important skill in data visualization literacy.
Reflecting on elementary school teaching, Chevalier \etal stressed the necessity of embedding critical thinking within visualization literacy education \cite{chevalier2018observations}.
Bergstrom and West, in their university course \cite{bergstrom2017calling}, gathered and deliberated on misleading instances of data visualization\leocw{}{s}, which were later compiled in their book \cite{bergstrom2021calling}.
In their hands-on teaching classroom, Lo \etal guided students to construct charts using visualization software and experience firsthand the inadequacy of the default axis direction in the context of ranking data, which \leocw{}{smaller value represents better ranking. This} inadvertently led to a downward \leocw{trending line for an}{trend for the} ascending ranking. Subsequently, the students were tasked to rectify the chart by adjusting the chart settings \cite{lo2019learning}.
Camba \etal investigated the effectiveness of various learning activities--such as in-class discussion\leocw{}{s}, self-learning, and peer challenge\leocw{}{s}--on learning to identify and accurately interpret misleading visualizations. They found that the peer challenge intervention\leocw{ resulted in}{} significantly improved performance in the post-intervention test. 
For this activity, students were asked to experiment with the dataset and different visualization settings\leocw{ in an attempt}{} to "deceive" their peers\leocw{,}{} and\leocw{ also to}{} identify deceptive techniques used in their peers' visualizations. This active learning approach substantially benefited students' learning outcomes \cite{camba2022identifying}.
}

\subsection{Automated Detection and Explanation of Misleading Visualizations}

The prevalence and impact of misleading visualizations in modern communication are significant, especially given the widespread availability of visualization tools. Automated detection and explanations serve as important countermeasures against these issues. Such detection tools are often referred to as linters, drawing an analogy to computer program code linters that raise warnings for constructs that may be legitimate but have the potential to cause errors. This analogy is fitting for detecting visualizations that are constructed correctly but may still be misleading.

McNutt and Kindlmann proposed a linter with linting rules derived from the Algebraic Visualization Design (AVD) framework \cite{mcnutt2018linting}.
They implemented a Python-based linter for charts created using the Python charting library\leocw{, }{--}matplotlib. In a course project,
Zheng and Sherif developed a JavaScript-based linter for the charting library, Chart.js \cite{zhengvisualization}.
Vizlinter proposed by Chen \etalcomma is built with Answer Set Programming (ASP), \leocw{}{and} it checks the visualization specifications of Vega-Lite \cite{chen2021vizlinter}.
These linters are rule-based, relying on predefined rules derived from visualization best practices.
Contrarily,\leocw{ in their work, Surfacing Visualization Mirage,}{} McNutt \etal applied the AVD framework to perform automatic checking on visualizations without predefined rules \cite{mcnutt2020surfacing}.

While the primary goal of these tools is to create automated checking programs that detect poorly constructed visualizations, their textual outputs \leocw{are often insufficiently}{need to be more} approachable for end users to understand the issue\leocw{}{s} and accept the \leocw{need for change}{needed changes}. These automatic detection tools provide only textual warnings, which may notify users but lack clear explanations of the problem and its location within the visualization.

Hopkins \etal proposed Visualint, a visual interface designed to highlight problematic regions of a chart and notify users \leocw{when}{that} the visualization is poorly constructed \cite{hopkins2020visualint}. Fan \etal suggested a pipeline that performs visualization reverse engineering \cite{poco2017reverse} to extract the visualization specification from its bitmap graphics form, then checks it against the rules derived from best practices \cite{fan2022annotating}. If any violations are detected, a small overlay is created on the chart for comparison with the suggested version.
While these tools have limitations, such as applicability to specific chart types, reliance on expert-created rules, and potentially high computational costs, they represent pioneering efforts in exploring the feasibility of creating user interfaces to help users understand issues and correct their visualizations.



\subsection{Explorable Explanations}

Bret Victor proposed the concept of \leocw{explorable explanations}{\expexp} as a powerful rhetorical device in writing \cite{victor2011explorable}. Enabled by web technology, interactive reading differs from traditional book reading by actively engaging readers with articles \leocw{that incorporate}{incorporating} \leocw{explorable explanations}{\expexp}. Instead of using static images and precalculated numbers, writers provide small widgets that allow readers to adjust \leocw{}{the} parameters and observe how the graphs or numbers change accordingly.

Dragicevic \etal introduced the idea of an explorable multiverse analysis report (EMAR) for reporting experimental results \cite{dragicevic2019increasing}. By adopting \leocw{explorable explanations}{\expexp}, readers can explore different settings of the experiment report, increasing transparency and mitigating issues related to reporting a single analysis path.

\leocw{Explorable explanations}{\expexp} \leocw{have}{has} significant potential for explaining various concepts, regardless of their complexity. A collection of \leocw{explorable explanations}{\expexp} is showcased on \href{https://explorabl.es/}{explorabl.es} \cite{Explorab74:online}, featuring numerous examples that cover complex mathematical and scientific concepts. One particularly inspiring example is Nicky Case's creative work on explaining the intricate concept of game theory\leocw{}{ in an interactive and engaging way with \expexp} \cite{case2017evolution}.\leocw{ It explains the complex concept of game theory in an interactive and engaging way.}{} We aim to incorporate \leocw{explorable explanations}{\expexp} into the explanation of visualization concepts, enabling readers to learn and be persuaded to identify misleading visualizations and accept suggested design changes more effectively.

\section{Existing Forms of Explanations}

In order to gain a thorough understanding of the common methods used to explain misleading visualizations, we collected examples from three distinct sources: the internet, books\leocw{}{,} and a design workshop. These examples illustrate why certain visualizations can be misleading and provide insight into the most effective ways to explain them.

\subsection{Examples from the Internet}

\begin{figure*}[t]
    \centering
    \includegraphics[width=\textwidth,keepaspectratio, trim={0 5cm 0 2.5cm},clip,page=1]{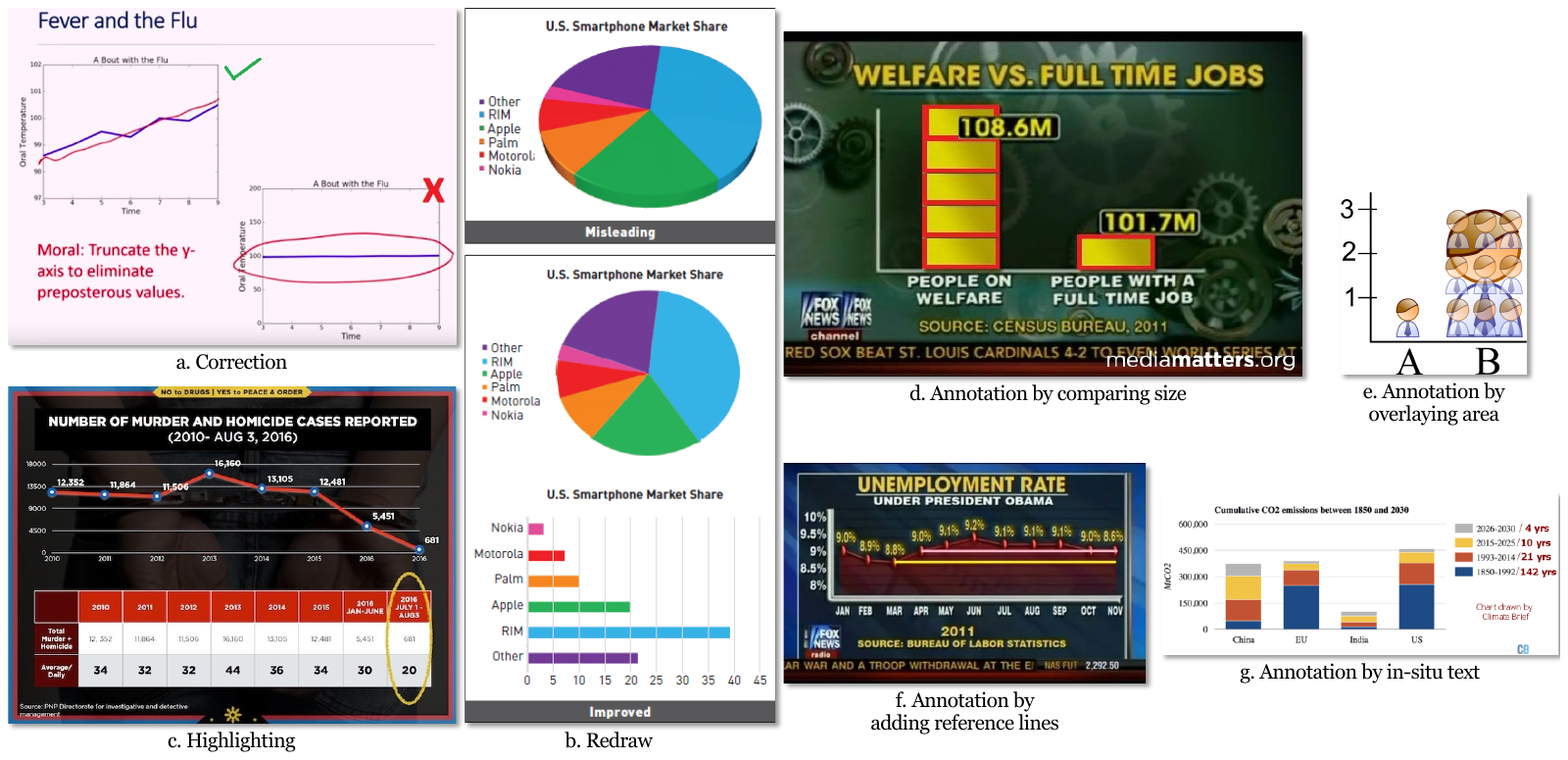}
    \squeezebetweenfigureandcaption
    \caption{Examples of visual \leocw{explanation collected}{explanations} from online \leocw{web sites}{websites} and social media. (a) Explaining the poor design choice of \leocw{}{the} y-axis range by providing the corrected version of the original chart. (b) Redrawing a 3D pie chart as a bar chart\leocw{,}{} or correction by removing the 3D effect. (c) Highlighting the inconsistent intervals of the last two x-axis elements in contrast to the other data points. (d) Annotating bar sizes by drawing boxes on top of the bars for easy comparison. (e) Filling the larger area with more small icons than the differences in data values suggest. (f) Drawing reference lines to emphasize the scale inconsistencies in the chart. (g) Adding in-situ text to point out the inconsistencies in bin sizes.}
    \label{fig:onlineexamples}
    \squeezeafterfigure
\end{figure*}

The internet offers an abundant resource for explanations concerning misleading visualizations. Numerous educational blog posts and social media discussions have been discovered in our research. We utilized the collection compiled by Lo \etal \cite{lo2022misinformed}, who assembled a dataset of improperly designed visualization examples sourced from search engines and social media platforms. Within this collection, 1,143 visualization examples are tagged with at least one issue. We leverage this dataset to investigate the methods employed by individuals online when explaining misleading visualizations to others.

We first attempted to access the web pages using the URLs provided in the dataset. As of January 2023, we successfully retrieved 992 documents out of the 1,143 entries. Some documents were irretrievable due to missing web pages or removal from social media platforms. Next, we filtered out duplicate entries with identical URLs and documents lacking explanations. Following the retrieval and filtering procedures, our study incorporated 360 documents containing explanations.
\leo{}{Notably, nearly a tenth ($9.4\%, N=34$) of these explanations were developed as instructional materials in the form of quizzes ($4.4\%,N=16$), slides ($3.9\%, N=14$), or course websites ($1.7\%, N=6$).}

Two authors coded these documents using the grounded theory method (GTM) as described by Muller \cite{muller2014curiosity}. GTM is a research approach for exploring an unfamiliar domain—in our case, understanding how to explain misleading visualizations. The coders iteratively analyzed the examples and discussed the explanation methods' definitions.
After iterations of coding and refining definitions, each tag achieved a Cohen's $\kappa > 0.7$. The coding results can be found in the supplemental materials.

The explanation methods fall into two categories: visual and textual, often complementing each other. Most explanations contain textual content ($93.9\%, N=338$). These explanations can be as concise as directly identifying issues, such as "Look at the percentages and the length of columns!"\footnote{\url{https://twitter.com/1GoldilocksZone/status/856973264526209024}}
We labeled these explanation methods as \leocw{short text}{\textbf{\shorttext}} ($48.3\%, N=174$). \leocw{Short text}{\shorttext} explanations are predominantly found on social media platforms like Twitter, where post length is limited. Conversely, \leocw{long text}{\textbf{\longtext}} explanations ($45.6\%, N=164$) are most commonly encountered in blog posts. They may involve in-depth discussions of a single issue\footnote{\url{https://andrewpwheeler.com/2013/08/28/hanging-rootograms-and-viz-differences-in-time-series/}} or address various visualization pitfalls using multiple examples\footnote{\url{https://flowingdata.com/2017/02/09/how-to-spot-visualization-lies/}}.

Over half of the explanations incorporate visual aids ($55.3\%, N=199$), with some even employing a combination of multiple techniques. The most prevalent visual aid contrasts two visualizations: a misleading version and a corrected or redrawn version. We differentiate between \leocw{correction}{\correction} and \leocw{redraw}{\redraw}, as \leocw{}{a} correction only modifies chart settings, while \leocw{}{a} redraw changes the entire chart type. For instance, changing the y-axis to an appropriate range to better show the trend in data is a correction (\cref{fig:onlineexamples}a), while converting a pie chart to a bar chart is a redraw (\cref{fig:onlineexamples}b). \leocw{Corrections}{\textbf{\correctionp}} ($41.7\%, N=150$) are more common than \leocw{redraws}{\textbf{\redrawp}} ($12.8\%, N=46$).

Another form of visual explanation involves marking the visualization to guide the audience in identifying the misleading elements in the chart. The simpler form is \leocw{highlighting}{\textbf{\highlighting}} ($5.8\%, N=21$), and the more complex form is \leocw{annotation}{\textbf{\annotation}} ($15\%, N=54$). \leocw{highlighting}{\highlighting} captures the audience's attention and emphasizes a specific area on the chart by circling the crucial part that leads to misunderstanding. This can also be achieved using arrows, a highlighter pen effect, or enlarging the region with a magnifying effect. \cref{fig:onlineexamples}c exemplifies circling a data point that creates a false impression of a decreasing trend. \leocw{Annotations}{\annotationp} draw the audience's attention and provide additional visual aids. The annotation in \cref{fig:onlineexamples}d simplifies the comparison between two bars for the audience. There are numerous other annotation forms, such as overlaying small icons on the larger icons in pictograms to facilitate area comparisons (\cref{fig:onlineexamples}e), adding reference lines to emphasize the scale inconsistencies in the chart (\cref{fig:onlineexamples}f), or explicitly marking chart elements with in-situ text to expose the inconsistent binning sizes (\cref{fig:onlineexamples}g).

\begin{table}
  \caption{Classification of six explanation methods identified from online explanation examples.}
  \centering
  \vspace{-10pt}
  \includegraphics[width=0.5\textwidth, clip, keepaspectratio, trim=1cm 21.5cm 7cm 0.5cm]{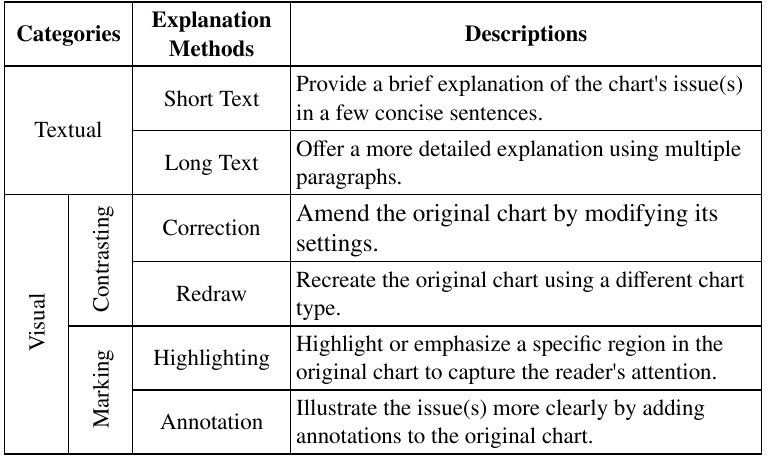}
  \label{table:explanationmethods}
  \squeezeafterfigure
  \vspace{-10pt}
\end{table}

\cref{table:explanationmethods} presents the categorization of explanation methods and their descriptions. In addition to these six identified methods, during the coding process, we also observed instances where evidence from various sources was compiled to form arguments that debunk claims derived from the charts. This evidence could include charts from other news sources, data from different data agencies, or other evidence like a prescription drug receipt to refute an incorrect insulin price chart. We conclude that although the categorization of these six explanation methods is not exhaustive in the design space of explaining misleading visualizations, it encompasses a representative set of explanations found on the internet, providing a foundation for those seeking to develop explanations for misleading visualizations.

\subsection{Explanations from Books}

The first documented discussion of misleading visualizations appears in Huff's book on statistics, which laid the groundwork for subsequent conversations on the subject. Books provide a valuable corpus for analysis, so we also examined them to learn how authors explain the examples they have collected. Through related book suggestions from Amazon and Goodreads, we gathered six titles: Huff \cite{huff1993lie}, Tufte \cite{tufte1983visual}, Monmonier \cite{monmonier2018lie}, Jones \cite{jones1995lie}, Cairo \cite{cairo2019charts}, and Bergstrom and West \cite{bergstrom2021calling}\leocw{}{.} These titles contain at least one chapter related to misleading visualizations, while other titles focus on identifying misleading uses of statistics but lack dedicated chapters on visualizations \cite{wheelan2013naked,stephens2018everybody,harford2022data,levitin2016field}.

We examined the chapters related to misleading charts and categorized the explanation methods employed by the authors. We found that contrasting techniques were used most frequently. The author would present a misleading visualization example encountered in daily life, guide readers in identifying the misleading elements in the chart, and then reveal the corrected or redrawn version as a truthful data representation. \leocw{annotation}{\annotation} is also a common technique\leocw{}{,} while \leocw{highlighting}{\highlighting} is the least common. These findings align with the results from the internet examples. We also noted Tufte's use of formulas to explain the Lie Factor, which is the only instance we found where a formula was used to \leocw{elucidate}{explain} why a chart is misleading.

\subsection{Design Workshop}

In addition to examples from the internet and books, we aimed to explore potential new explanation methods. We organized a design workshop with 16 participants, consisting of postgraduate students and researchers in the field of data visualization research.

The workshop comprised three parts. In the first part, participants answered six questions related to various chart issues. The initial five questions each included two visualizations—one misleading and one corrected version. Participants were asked to explain why one chart was preferable to the other and to convey their explanations to a layperson from the general public. They were encouraged to use non-textual aids in their explanations. The fifth question was similar to the first four but lacked a corrected version. The final question asked participants to justify their explanations. In this part, we aimed to gather ideas about explanation methods beyond those collected from the internet and books.

The second part introduced techniques found in the \leocw{Internet}{internet} examples to participants, aiming to provide hints or spark ideas for enhancing their initial answers. In the third part, participants revisited their explanations and attempted to improve them using techniques introduced in the second part. Lastly, they were asked to share their thoughts on the changes. This part aimed to explore the benefits of providing explanation techniques and examples for those forming explanations and rationalizing visualization design choices.

From the design workshop, we identified three new variants of explanation techniques: (1) \leocw{analogy}{Analogy}, (2) \leocw{breakdown}{Breakdown}, and (3) \leocw{animation}{Animation}. Analogy is akin to Huff's chart background explanations for truncated axes \leocw{}{\cite{huff1993lie}}. Breakdown shows the decomposition of the largest group in the space next to the original chart, such as the largest group in a histogram. Animation involves rotating or morphing a chart into its correct form. As \leocw{analogy}{Analogy} and \leocw{breakdown}{Breakdown} are drawn on the original chart to assist explanation, they may be categorized as \leocw{annotations}{\annotationp}. Animation is similar to contrasting techniques like \leocw{correction}{\correction} and \leocw{redraw}{\redraw} but transforms the original chart into the corrected or redrawn version\leocw{}{ with animation}. The workshop materials and coding results can be found in the supplemental materials.

Over three-quarters of the participants ($75\%, N=12$) enhanced their explanations in part three. Participants who initially relied on textual explanations were able to enrich their explanations with various visual aids across different questions. \leocw{highlighting}{\highlighting} was the most frequently used technique ($52.5\%, N=42$) in part one.

\leo{Through this formative study of explanation methods}{Through the explanations collected from the internet, books, and design workshop}, we identified six major forms: \leocw{(1) correction, (2) redraw, (3) highlighting, (4) annotation, (5) short text, and (6) long text.}{(1) \shorttext, (2) \longtext, (3) \correction, (4) \redraw, (5) \highlighting, and (6) \annotation.} This classification informs the design of explanations \leocw{used to}{that} facilitate learning and persuade visualization tool users to make better design choices. All the collected data and materials are available on OSF\footnote{\osf\label{osf}}.

\begin{figure*}[th]
    \centering
    \includegraphics[width=\textwidth,keepaspectratio, page=2,trim={0 8.4cm 0 2.5cm},clip]{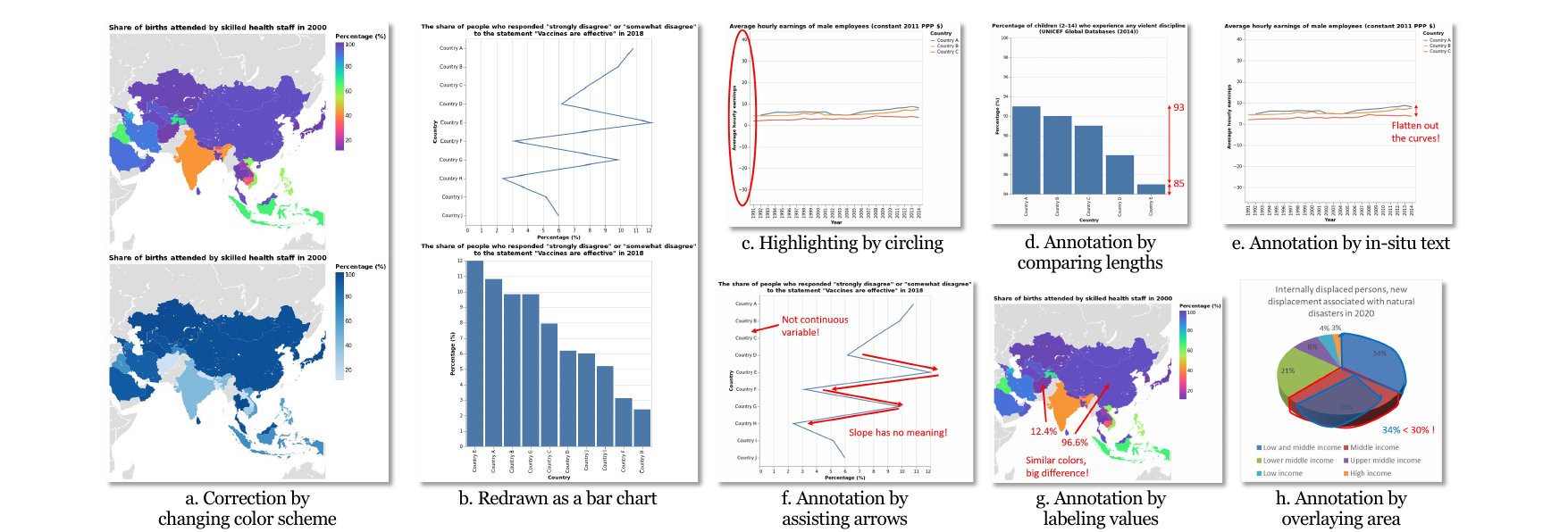}
    \caption{Examples of implemented visual explanation methods. (a) Correcting a rainbow color choropleth to a sequential color scheme. (b) Redrawing a horizontal line chart with categorical variables into a bar chart. (c) Highlighting the improperly configured axis that downplay\leocw{}{s} the data variations. (d) Using lines to annotate the mismatch between bar lengths and data values in the chart. (e) Annotating to emphasize the impact of \leocw{}{an} improperly configured axis that minimize\leocw{}{s} data fluctuations. (f) Using arrows to emphasize the irrelevance of sharp inclines in a line chart with categorical variables. (g) Pointing out the confusion caused by distinct values being represented by similar hues in a rainbow color scheme. (h) Overlaying the bottom pie chart slice with the top pie chart slice that has a downplayed area caused by the 3D perspective effect.}
    \label{fig:implementation}
    \squeezeafterfigure
\end{figure*}

\section{Explanation Designs}

For the evaluation study, we selected five issues from the taxonomy \leocw{}{proposed} by Lo \etal \cite{lo2022misinformed} and five explanation methods from the formative study. Apart from the implementation for evaluation purposes, we also examined the feasibility and challenges of applying different explanation methods.

\subsection{Construction of Misleading Visualizations}
\label{section:constructionmisleadingvis}

\leo{We chose five issue types from the taxonomy by Lo \etal \mbox{\cite{lo2022misinformed}}, which were categorized under the design and perception stages. These two stages are most relevant to chart design choices and how people perceive visualizations.}{
In selecting visualization issues to demonstrate the explanation methods, we aimed to include a diverse range of common design-related visualization pitfalls and different chart types.
\leocw{From the five-stage taxonomy proposed by Lo \etal \mbox{\cite{lo2022misinformed}}, the design and perception stages are the most relevant.}{The design and perception stages are the most relevant from the five-stage taxonomy proposed by Lo \etal \cite{lo2022misinformed}.}}
From these two stages, we selected three design choice issues and two perception issues \leo{}{\leocw{that have}{with} the highest occurrences in their corresponding categories}. This selection covers the choice of axes, chart types, color schemes, and the design choice between 2D and 3D.
\leo{The issues we chose were (1) truncated axis on bar chart, (2) inappropriate axis range on line chart, (3) inappropriate use of line chart, (4) ineffective color scheme on choropleth, and (5) 3D pie chart.}{We selected issues that comprise:
(1) \textbf{Truncated Axis} on bar charts, where the differences between bar lengths are exaggerated due to a non-zero starting point on the y-axis. \cref{fig:implementation}d provides an annotated explanation of this issue.
(2) \textbf{Inappropriate Axis Range} on line charts, where the lines appear predominantly flat due to the expansive range of the y-axis (\cref{fig:implementation}e).
(3) \textbf{Inappropriate Use of Line Chart}, where a categorical variable is erroneously encoded on the y-axis (\cref{fig:implementation}f).
(4) \textbf{Ineffective Color Scheme} on choropleth maps, where a rainbow color scheme is used to represent a continuous variable (\cref{fig:implementation}g).
(5) \textbf{3D} pie chart, where the pie chart is presented with a perspective distortion (\cref{fig:implementation}h).}
These issues span the most common chart types--bar charts, line charts, pie charts, and choropleths--and highlight their most prevalent issues.

In the explanation examples collected in Section 3, some explanations use charts constructed from synthetic data to better illustrate the problem. However, synthetic data may risk influencing experiment results by causing participants to question the data instead of the chart design. To avoid this, we used real data to construct the charts and clearly indicated the data sources \leocw{at the bottom}{next to the chart}. We randomly picked datasets from the list of all available charts on Our World In Data \cite{ChartsOu50:online}\leocw{}{.} To avoid geographical biases, country names were masked as Country A, B, C, etc., except \leocw{for}{in} choropleth \leocw{,}{maps.} \leocw{and the}{The} same practice applied to continents.

We employed Python libraries pandas \cite{mckinney2011pandas} and Altair \cite{vanderplas2018altair} within the Jupyter Lab environment to construct the charts, except for 3D pie charts, which we created using Microsoft Excel with VBA scripts due to a lack of available Python libraries for 3D pie charts.
\leo{We constructed a set of 60 charts, with 30 being misleading and 30 not misleading.}{We constructed a total of 60 charts, which were then randomly divided into two sets.
The division criteria ensured that each of the five issues was represented by three misleading charts and their respective corrected versions, thus resulting in $5 \times 3 \times 2 = 30$ charts in each set.
In \leocw{experiment one}{Experiment One}, the pre-intervention test used the first set, while the post-intervention test used the second set.
In \leocw{experiment two}{Experiment Two}, only the first set is used.
More specifics on the experimental design are provided in \cref{section:experiment1} and \cref{section:experiment2}.}
The datasets and scripts to construct these charts are available on OSF\footref{osf}.

\begin{figure*}[th]
    \centering
    \includegraphics[width=\textwidth,page=3,keepaspectratio, trim={0 5.5cm 0 3cm},clip]{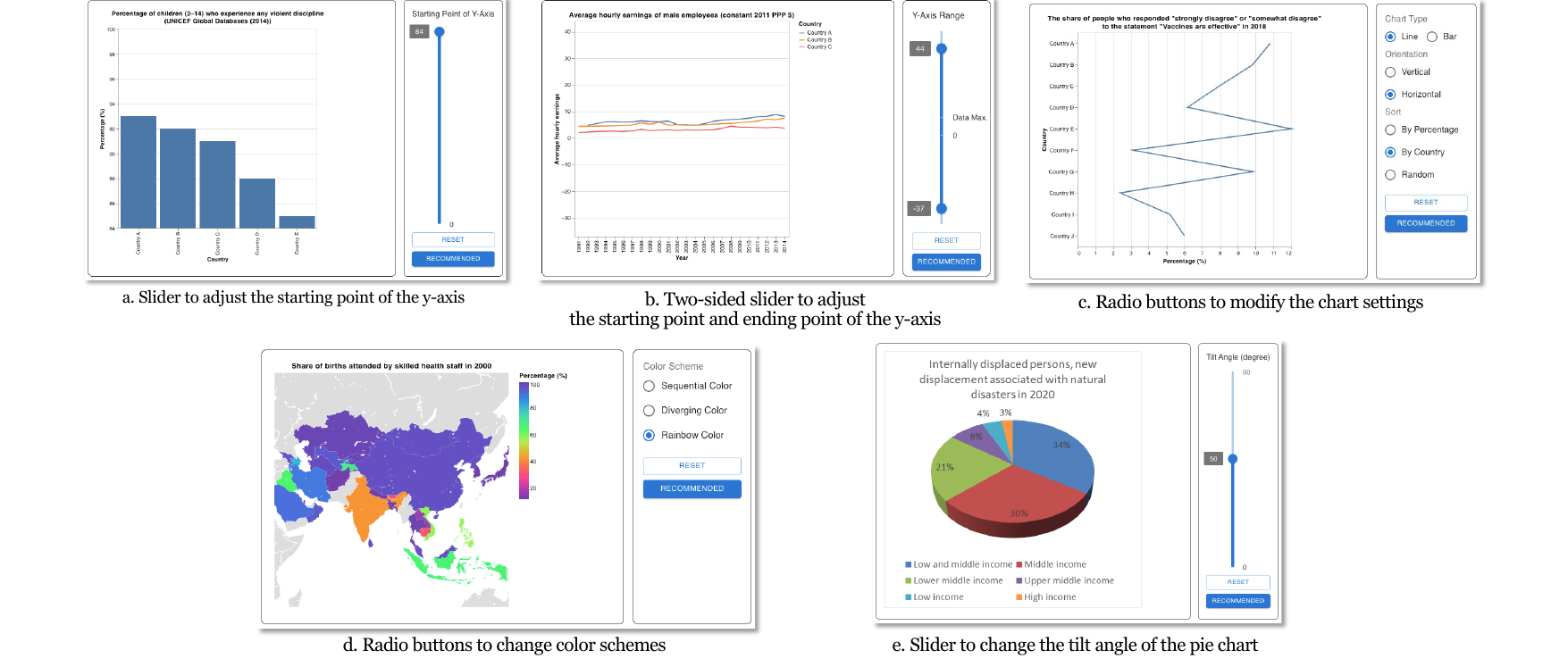}
    \caption{
    Examples of implemented \leocw{explorable explanations}{\expexp}. (a) By adjusting the slider, readers can experiment with different values of the y-axis starting point to see the changes \leocw{of}{in} the bar lengths. (b) Readers can set the y-axis range by adjusting the two-sided slider to observe the changes \leocw{of}{in the} line chart. (c) The radio buttons on the right let readers experiment with different chart settings. (d) Readers can choose different color schemes by clicking the radio buttons. (e) Readers can change the tilt angle of the pie chart to see the changes \leocw{of}{in the} slice area.}
    \label{fig:explorables}
    \squeezeafterfigure
\end{figure*}

\subsection{Explanation Methods}

We implemented five designs to explain the above issues: (1) \leocw{short text}{\shorttext}, (2) \leocw{correction}{\correction} or \leocw{redraw}{\redraw}, (3) \leocw{highlighting}{\highlighting}, (4) \leocw{annotation}{\annotation}, and (5) \leocw{explorable explanations}{\expexp}. \leocw{Short text}{\textbf{\shorttext}}, the simplest and most commonly used explanation \leocw{}{method}, sets the baseline for our evaluation experiment. \leocw{Second, correction}{\textbf{\correction}} and \leocw{redraw}{\textbf{\redraw}} both contrast the corrected version and the original misleading version, differing in whether the chart type is changed. \leocw{Of the five issues we studied, four can be fixed by correction without changing the chart type. In contrast, the inappropriate use of line charts}{Four of the five issues we studied can be corrected without changing the chart type, with the only exception of \leocw{inappropriate use of line charts}{Inappropriate Use of Line Chart}, which} requires a redraw of the line \leocw{charts}{chart} to \leocw{}{a} bar \leocw{charts}{chart} \leocw{from avoiding}{to avoid} misleading encoding of categorical variables. \leocw{Highlighting}{\textbf{\highlighting}} can be done by circling chart elements related to the issues, such as the starting point of the y-axis or the legend of rainbow colors. Implementing \leocw{annotation}{\textbf{\annotation}} requires more sophisticated thinking to convey and explain the message to readers. We learned from the collected examples from the \leocw{Internet}{internet} and books that applied the \leocw{annotation}{\annotation} technique, which includes using lines, arrows, short phrases, and overlapping areas to compare sizes. Examples of these designs can be seen in  \cref{fig:implementation}.

\leocw{Explorable explanation}{\textbf{\expexp}} is a concept proposed by Bret Victor \cite{victor2011explorable}. It allows readers to understand the content and interact with and challenge the writer's claim. It provides transparency, enables exploration, builds confidence and trust in the arguments, and encourages readers to form their own ideas and actively test them against the writer's message. Throughout the readings, there are controllable widgets that allow readers to tweak and see the effect on the numbers or graphics. We hypothesize that \leocw{explorable explanations}{\expexp} \leocw{are}{is} a good match for explaining misleading visualizations. They enable readers not only to accept visualization best practices but also to explore different parameter settings and see how they affect the charts.

\leocw{In the five issues, we implemented explorable explanations through sliders and radio buttons}{We implemented explorable explanations through sliders and radio buttons for the five issues}, enabling readers to interact with different parameter settings and see their effects on the charts. For \leocw{truncated axis}{Truncated Axis} and \leocw{inappropriate axis range}{Inappropriate Axis Range}, readers can move sliders to change the axis start\leocw{ing}{} or end\leocw{ing}{} points. The same sliders apply to 3D pie charts, allowing readers to change the perspective angle. For choropleths, readers can change the color scheme between sequential, diverging, and rainbow colors. For \leocw{inappropriate use of line charts}{Inappropriate Use of Line Chart}, readers can configure the line chart with different item ordering, orientation, and chart types between line \leocw{charts}{chart} and bar \leocw{charts}{chart}. \cref{fig:explorables} shows examples of\leocw{ the}{} implemented \leocw{explorable explanations}{\expexp}.

\section{Evaluation Experiments}
\label{section:experiments}

We aim to evaluate the effectiveness of different explanation methods in two aspects: (1) learning to identify visualizations that violate visualization guidelines, and (2) persuading visualization tool users to choose a design that follows the visualization guidelines. We designed two experiments and conducted them on prolific.ac, a crowdsourcing platform focused on studies rather than the more general crowdsourcing platform provided by Amazon. The recommended hourly rate on the Prolific platform is GBP 9 (\textasciitilde USD 11).

For both studies, we require participants to be fluent in English, have no colorblindness, and maintain an approval rate of 98\% or higher. On the Prolific platform, 75,552 workers meet these criteria, accounting for $62.2\%$ of the total 121,407 available workers. The criterion of being free from colorblindness ($70.4\%, N=85,445$) is the major factor reducing the suitable worker pool. \leocw{However}{Despite that}, this criterion is necessary to exclude variations that may affect the results related to color scheme design choices. Besides these criteria, we do not impose further restrictions on gender, age, geographic location, or education level.

Both studies begin with a consent form, a background information form, and a subjective literacy assessment form \cite{garcia2016measuring}. The subjective graph literacy (SGL) assessment consists of ten questions, each asking participants to rate their level of competence in chart reading and reliance on graphical information on a six-point scale. The assessment is empirically tested to have a strong positive correlation with the objective graph literacy (OGL) \cite{galesic2011graph} score.

\subsection{Experiment One: Learning Effects of Explanations}
\label{section:experiment1}

We are interested in testing the effectiveness of explanations in educating readers to identify charts that violate visualization guidelines. In this experiment, we have set up five conditions: (1) \leocw{short text}{\shorttext}, (2) \leocw{short text}{\shorttext} + \leocw{highlighting}{\highlighting}, (3) \leocw{short text}{\shorttext} + \leocw{annotation}{\annotation}, (4) \leocw{short text}{\shorttext} + \leocw{correction}{\correction}, and (5) \leocw{short text}{\shorttext} + \leocw{correction}{\correction} + \leocw{explorable explanations}{\expexp}. The \leocw{short text}{\shorttext} condition serves as a baseline for comparison with the \leocw{highlighting}{\highlighting}, \leocw{annotation}{\annotation}, and \leocw{correction}{\correction} conditions. Condition 5 requires combining \leocw{explorable explanations}{\expexp} with \leocw{correction}{\correction} because the corrected version of the chart \leocw{will}{is used to} guide participants to explore the explanation.

\subsubsection{Experiment Procedures}

\begin{figure}
    \centering
    \includegraphics[width=0.475\textwidth,keepaspectratio]{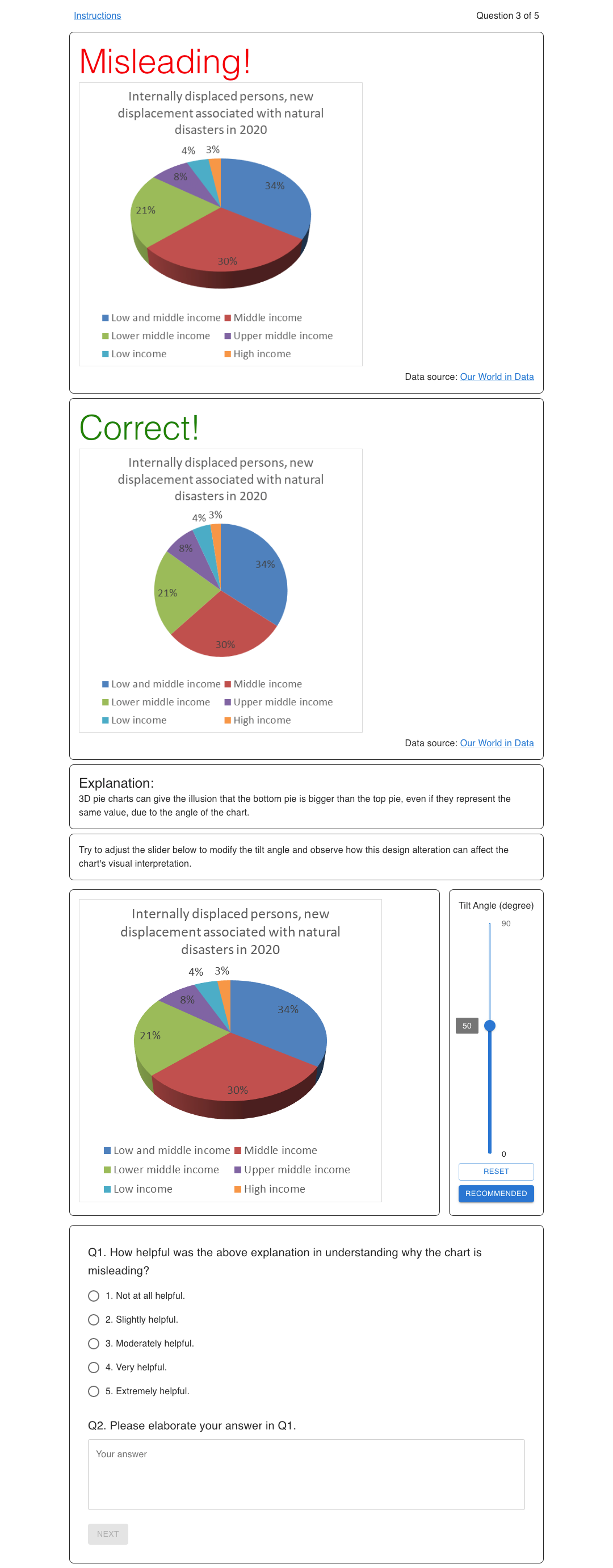}
    \caption{Screenshot of \leocw{experiment one}{Experiment One} Condition 5 (\leocw{short text}{\shorttext} + \leocw{correction}{\correction} + Explorable \leocw{Explanations}{Explanation}). The other conditions are similar to this interface. Condition 1 has only the text explanation. Condition\leocw{}{s} 2 and 3 have the text explanations and \leocw{highlighting}{\highlighting} or \leocw{annotation}{\annotation} but not the correction. Condition 4 \leocw{have}{has} the same interface but not the explorable view.}
    \label{fig:exp1interface}
\end{figure}

The study is conducted using a between-subject design \leocw{with}{in} three phases, including a pre-intervention testing phase, an intervention phase, and a post-intervention testing phase.

After collecting \leocw{}{the} basic information and completing the SGL assessment, participants will first perform a test to gauge their ability to distinguish between misleading visualizations and those that accurately represent data.
\leo{The test consists of 30 visualizations. For each of the five issues, three misleading charts are constructed with datasets randomly picked from Our World in Data, and three corrected charts are also constructed, resulting in $5 \times 3 \times 2 = 30$ charts.}{The test \leocw{consists of}{comprises} 30 visualizations, including three misleading charts for each of the five issues and their respective corrected versions, \ie, the first set described in \cref{section:constructionmisleadingvis}.}
The order of the charts is randomized across participants. Participants answer the question with one of the options "misleading," "accurate," or "I am not sure."
The screenshots of a sample test are included in the supplemental materials and are available on OSF\footref{osf}.

In the intervention phase, participants view\leocw{ed}{} a series of five explanations constructed according to their assigned conditions. Participants are asked to rate the helpfulness of the explanation and elaborate on their rating. \cref{fig:exp1interface} shows the screenshot of the \leocw{interface that the participants saw}{participants' interface} in Condition 5. The order of the explanations on different issues is randomized across participants. After reading the explanations, participants are required to complete the post-intervention test, which consists of \leo{another set of 30 visualizations constructed similarly to the pre-intervention test.}{a different set of 30 visualizations with a similar composition \leocw{as}{to} the pre-intervention test, \ie, the second set described in \cref{section:constructionmisleadingvis}.} Upon completion, participants are thanked and given an opportunity to report any comments or issues they encountered during the experiment.

\subsubsection{Hypotheses}

In Experiment \leocw{one}{One}, we want to test the effectiveness of explanations on participants' ability to identify charts that violate visualization guidelines. We hypothesize that the intervention, \ie, exposure to the explanations, has an effect on participants' ability to distinguish misleading charts. Therefore, our first hypothesis is: \textbf{Hypothesis 1.1 The correctness in identifying misleading charts and accurate charts is higher in the post-intervention phase, regardless of \leocw{}{the} explanation method\leocw{s}{}.}

\leo{Secondly, we believe some explanation methods can better explain why a visualization is misleading and affect participants' performance in the post-intervention phase.}{
Secondly, \leocw{explorable explanations}{\expexp} \leocw{have}{has} demonstrated \leocw{the}{its} effectiveness in illustrating the linkage between variable parameters and their impact on outcomes.
In the context of visualizations, these parameters involve\leocw{ the}{} design decisions such as the start\leocw{ing}{} and end\leocw{ing}{} points of the axis, or the color \leocw{schemes}{scheme} applied in the chart.
We posit that \leocw{explorable explanations}{\expexp} can also be used effectively to explain visualization guidelines.} Hence, our second hypothesis is: \textbf{Hypothesis 1.2 \leo{The correctness in pre-intervention and post-intervention phases across different conditions exhibits varying degrees of change.}{The \leocw{explorable explanations}{\expexp} condition has a greater effect on improving the correctness in the post-intervention phase.}}

\leo{Thirdly, despite the significant efforts of the visualization community in studying the inaccurate perception of rainbow colors and 3D pie charts and advocating for better design choices of sequential colors and 2D pie charts, people still prefer rainbow colors and 3D pie charts. We speculate that this preference is hard to change, and participants may not consider them misleading. Our third hypothesis is: Hypothesis 1.3 The correctness in pre-intervention and post-intervention phases across different issue types exhibits varying degrees of change.}{}

\subsubsection{Participants}

For each condition, we recruited 25 participants on the Prolific platform according to the criteria described at the beginning of \leocw{Section 5}{\cref{section:experiments}}. In total, 125 participants took part in the experiment. Among the participants, 57.6\% identified as male, 40.8\% as female, and 1.6\% as other. The reported age ranged from 19 to 54 ($age_{median}: 24, age_{mean}: 26.2, age_{std}: 6.06$). The reported highest completed education levels were Bachelor's degree (41.6\%), some college credit (23.2\%), high school graduation (21\%), Master's degree (12.8\%), and Doctorate's degree (1.6\%). The most common occupation was student ($40\%, N=50$); the rest were mostly white\leocw{ }{-}collar workers, except a cook\leocw{}{ing} assistant, receptionist, warehouse worker, dentist, and military personnel. The participants' nationalities were mainly from Europe (67.2\%), with the remainder from Africa (18.4\%), North America (9.6\%), South America (2.4\%), and Asia (2.4\%). The SGL assessment (10-item on a six-level scale) had a median of 4.4 ($SGL_{mean}: 4.32, SGL_{std}: 0.86, SGL_{min}: 2.5, SGL_{max}: 5.9$). The median completion time for Experiment \leocw{one}{One} was 23.78 minutes. The study was advertised to workers as a 25-minute study with a compensation of GBP 3.75 (\textasciitilde USD 4.63), translating to an effective hourly rate of GBP 9 (\textasciitilde USD 11). Two participants suspected of answering randomly ($15 \pm 1$ out of 30 correct in both prior\leocw{}{-} and post\leo{}{-}intervention tests) were checked, and their submissions were removed from the study.

\subsubsection{Results}

\begin{figure}
    \centering
    \includegraphics[width=0.5\textwidth,keepaspectratio, trim={0 22cm 5cm 0},clip]{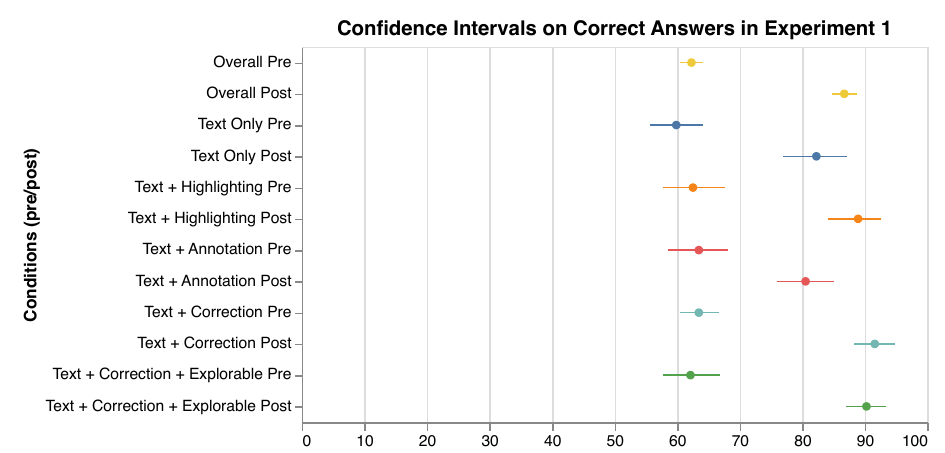}
    \squeezebetweenfigureandcaption
    \caption{Confidence intervals of correct answers in the prior\leocw{}{-} and post\leocw{ }{-}intervention tests. Participants \leocw{had perform}{performed} significantly better after reading any of the explanations. However, we did not observe significant differences across conditions.}
    \label{fig:exp1ciplot}
    \squeezeafterfigure
\end{figure}

The median number of correctly answered questions on the pre-intervention test was 19 out of 30 ($63.3\%, correct_{mean}: 18.69, correct_{std}: 3.38, correct_{min}: 11, correct_{max}: 27, N=125$), while the post-intervention median was 27 ($90\%, correct_{mean}: 26.02, correct_{std}: 3.44, correct_{min}: 16, correct_{max}: 30, N=125$). An ANOVA test comparing the two measures rejected the null hypothesis for hypothesis 1.1, indicating a significant positive change in participants' ability to correctly identify misleading and accurate charts in the post-intervention test ($F(1,248) = 288, p < 0.0001$). \cref{fig:exp1ciplot} shows the confidence interval plots for each condition and the overall results.

Additionally, we observed the median precision, recall, and F1 scores for the pre-intervention phase as $0.71$, $0.53$, and $0.61$, respectively, and the post-intervention phase as $0.92$, $0.93$, and $0.91$. In the pre-intervention test, despite instructions specifying a $15:15$ ratio between misleading and accurate visualizations, participants were more likely to identify charts as accurate rather than misleading, with a ratio of $10.8:17.4$, suggesting they only labeled charts as misleading when confident in their decision. This ratio shifted to $15.1:14.4$ in the post-intervention test.

\leo{We conducted post hoc pairwise t-tests with Bonferroni correction across different conditions. Condition 3 (Text + \leocw{annotation}{\annotation}) showed significantly lower improvement compared to other conditions with visual explanations: vs. Condition 2 (Text + \leocw{highlighting}{\highlighting}, $p: 0.066$), vs. Condition 4 (Text + \leocw{correction}{\correction}, $p: 0.020$), and vs. Condition 5 (Text + \leocw{correction}{\correction} + \leocw{explorable explanation}{\expexp}, $p: 0.008$), but not against the baseline Condition 1 (Text Only). No significant differences were found among other condition pairs ($p > 0.1$). Although these results reject the null hypothesis for hypothesis 1.2, indicating significant differences among all conditions, interpreting the findings is challenging. It is counterintuitive that visual explanations did not yield significant benefits over text explanations alone.}{
We conducted post hoc pairwise t-tests with Bonferroni correction between Condition 5 (Text + \leocw{correction}{\correction} + \leocw{explorable explanation}{\expexp}) and other conditions.
There was a significant difference when compared to Condition 3 (Text + \leocw{annotation}{\annotation}, $p: 0.008$).
However, no significant differences were observed when \leocw{comparing}{compared} with other conditions.
Therefore, the null hypothesis of hypothesis 1.2 is not rejected.

Aside from Condition 5, it is worth noting that Condition 3 (Text + \leocw{annotation}{\annotation}) showed significantly lower improvement compared to other conditions with visual explanations: vs. Condition 2 (Text + \leocw{highlighting}{\highlighting}, $p: 0.066$), vs. Condition 4 (Text + \leocw{correction}{\correction}, $p: 0.020$), and vs. Condition 5 (Text + \leocw{correction}{\correction} + \leocw{explorable explanation}{\expexp}, $p: 0.008$).
\leocw{However, when compared to the baseline Condition 1 (Text Only), no significant difference was found ($p: 1.0$).}{Despite that, no significant difference was found when compared to baseline Condition 1 (Text Only, $p: 1.0$).}}

\leo{For each issue type, we also performed post hoc pairwise t-tests with Bonferroni correction. Among the five pairwise tests, only Condition 5 (Text + Correction + Explorable Explanation) with 3D pie charts showed significant differences compared to other conditions: vs. Condition 1 ($p: 0.098$), vs. Condition 2 (p: 0.005), vs. Condition 3 ($p: < 0.001$), except for Condition 4 (Text + \leocw{correction}{\correction}). Additionally, Condition 4 significantly differed from Condition 3 ($p: 0.005$) but not from other conditions. Regarding the issue of rainbow color, no significant differences were found between methods, except for Condition 5 vs. Condition 3 ($p: 0.037$). While we detected significant differences across conditions for various issue types, these results may be prone to p hacking, as there were 50 pairwise comparisons and only five significant findings ($p < 0.1$).}{}


\subsection{Experiment Two: Persuasive Effects of Explanations}
\label{section:experiment2}

One of the goals of this study is to inform the future development of visualization tools by suggesting chart fixes that align with visualization guidelines. We designed \leocw{experiment two}{Experiment Two} to simulate a situation where participants are persuaded to abandon their choice of a misleading chart and switch to one that aligns with visualization guidelines. In this experiment, we established five conditions: (1) \leocw{correction}{\correction} only, (2) \leocw{correction}{\correction} + \leocw{short text}{\shorttext}, (3) \leocw{correction}{\correction} + \leocw{short text}{\shorttext} + \leocw{highlighting}{\highlighting}, (4) \leocw{correction}{\correction} + \leocw{short text}{\shorttext} + \leocw{annotation}{\annotation}, and (5) \leocw{correction}{\correction} + \leocw{short text}{\shorttext} + \leocw{explorable explanations}{\expexp}. Since we suggest a fix to users, the base condition shows only the corrected version. Condition 2 adds on \leocw{}{the base} condition with \leocw{short text}{\shorttext} explanations. Conditions 3, 4, and 5 provide further visual explanations.

\subsubsection{Experiment Procedures}

\begin{figure}
    \centering
    \includegraphics[width=0.475\textwidth,keepaspectratio]{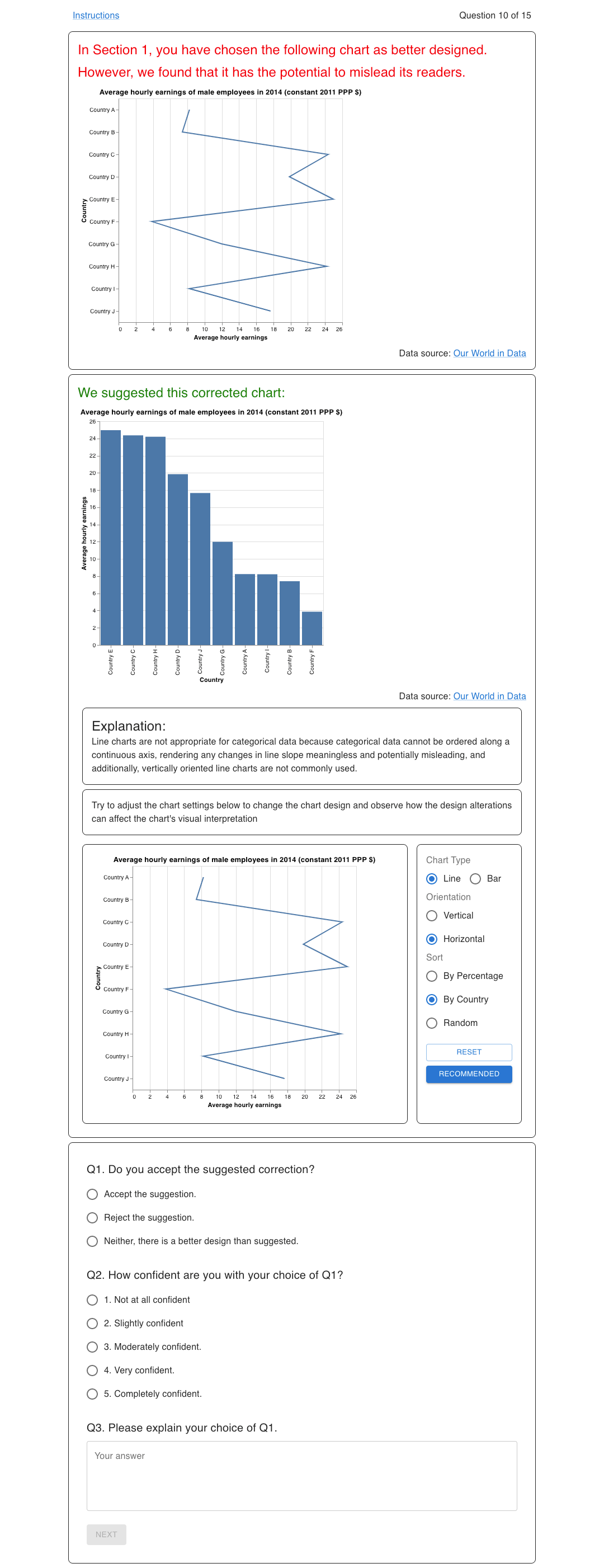}
    \caption{Screenshot of \leocw{experiment two}{Experiment Two} Condition 5 (\leocw{correction}{\correction} + \leocw{short text}{\shorttext} + \expexp). The other conditions are similar to this interface. Condition 1 has only the correction. Condition 2 has the correction and the text explanations. Condition\leocw{}{s} 3 and 4 have the same interface\leocw{ by}{,} replacing the explorable view with \leocw{highlighting}{\highlighting} or \leocw{annotation}{\annotation}.}
    \label{fig:exp2interface}
\end{figure}

Experiment \leocw{two}{Two} employs a between-subject design with only two phases. \leocw{As in the first experiment}{Same as Experiment One}, participants begin by completing basic information and \leocw{}{an} SGL assessment. The first phase includes 15 this-or-that questions, asking participants to choose one of two presented chart designs as the better one. One chart in each pair exhibits one of the five misleading issues, while the other is a corrected version. Each issue has three pairs of charts in the question set. The visualization pairs and their order within the pair are randomized.

Based on answers from the first phase, the chosen misleading charts are presented again in the second phase, accompanied by explanations of why the selected charts are misleading and why the alternative charts are better. Participants are asked to accept or reject the suggestion\leocw{}{,} or choose neither \leocw{option}{options}. They must also explain their choice and rate their confidence on a five-option Likert scale.
\cref{fig:exp2interface} shows the screenshot of the interface \leocw{that the participants saw}{displayed to the participants} in Condition 5. After completing the experiment, participants are thanked and allowed to report any comments or issues they encounter\leocw{}{ed}.

\subsubsection{Hypotheses}

\leocw{In experiment two, we aim to test if participants might change their minds and accept the suggested chart version.}{Experiment Two aims to test whether participants might change their minds and accept the suggested chart version.} We hypothesize that providing an explanation alongside the corrected chart is more convincing than presenting only the corrected chart. Thus, \leocw{}{our first hypothesis is:} \textbf{Hypothesis 2.1 \leocw{states that}{The} participants in conditions with more than just the corrected chart will have a higher acceptance rate for the suggested chart.}

Secondly, similar to \leocw{experiment one}{Experiment One}, we believe that people's strong preferences for rainbow colors and 3D pie charts make it difficult to persuade them to choose less visually appealing monotonic sequential colors and 2D pie charts. Our second hypothesis is: \textbf{Hypothesis 2.2 \leocw{posits that}{The} acceptance rates across different issue types will exhibit significant differences regardless of explanation methods.}

Lastly, we have observed the effectiveness of \leocw{explorable explanations}{\expexp} in clarifying complex concepts across various subjects, from mathematics to social sciences. We contend that applying \leocw{explorable explanations}{\expexp} will help communicate visualization guidelines to participants and persuade them to accept the suggested chart design. Our third hypothesis is: \textbf{Hypothesis 2.3 \leocw{asserts that the}{The} acceptance rate in the \leocw{explorable explanation}{\expexp} condition will be higher than in other conditions across different issue types.}

\subsubsection{Participants}

For each of the five conditions, we recruited 25 participants through the Prolific platform based on the criteria described at the beginning of \leocw{Section 5}{\cref{section:experiments}}. Participants who participated in \leocw{experiment one}{Experiment One} were excluded from the worker pool of \leocw{experiment two}{Experiment Two}. In total, 125 participants took part in the experiment.

Participants in the second experiment were 48.8\% male, 50.4\% female, and 0.8\% other\leocw{. Ages}{, and their age} ranged from 18 to 55 ($age_{median}: 25, age_{mean}: 27.2, age_{std}: 7.78$). \leocw{Reported}{The reported} highest completed education levels included Bachelor's degree (47.2\%), some college credit (24\%), high school graduation (14.4\%), Master's degree (11.2\%), and Doctorate's degree (3.2\%). The most common occupation was student (31\%, N=38), with the remainder mostly \leocw{consisting of}{comprising} white\leocw{ }{-}collar workers, with some exceptions such as homemakers and bus drivers. Participants' nationalities were mainly \leocw{from Europe}{European} (50.4\%) and \leocw{Africa}{African} (28.8\%), with the \leocw{rest from}{remainder being} North \leocw{America}{American} (16.8\%), \leocw{Asia}{Asian} (2.4\%), South \leocw{America}{American} (0.8\%), and \leocw{Oceania}{Oceanic} (0.8\%). The SGL assessment (10-item on a six-level scale) \leocw{had}{showed} a median of 4.5 ($SGL_{mean}: 4.4, SGL_{std}: 0.81, SGL_{min}: 2.0, SGL_{max}: 6.0$). The median completion time for \leocw{experiment two}{Experiment Two} was 13.57 minutes. The study was advertised as a 15-minute study with a GBP 2.25 (\textasciitilde USD 2.78) compensation, yielding an effective hourly rate of GBP 9 (\textasciitilde USD 11).

\subsubsection{Results}

The median number of correctly answered questions in the first phase was 11 out of 15 ($73.3\%, correct_{mean}: 10.66, correct_{std}: 2.68, correct_{min}: 4, correct_{max}: 15, N=125$). This result is higher than the median in \leocw{experiment one}{Experiment One}, which was 63.3\%. When presented with two versions of the charts, participants were able to\leocw{ better}{} distinguish the misleading chart \leocw{}{better} and choose the correct one, as shown by the ANOVA test ($F(1,248)=21.67, p < 0.001$).

Since the number of incorrect answers in phase one varied among participants, the number of visualization pairs presented to each participant also differed. The same applied to issue types. While 542 explanations were presented across all participants, only 17 were related to \leocw{inappropriate axis range}{Inappropriate Axis Range}, and 26 were concerned with \leocw{inappropriate use of line charts}{Inappropriate Use of Line Chart}. Due to the limited sample size, we excluded these two issues from the following analysis.

We calculated the acceptance rate as $\frac{A}{S}$, where A is the number of accepted suggestions, and S is the number of suggestions presented to the participant. Participants who received no suggestions, \ie, those who answered correctly on all nine questions, were not counted. In total, 17 participants scored full marks: 4 for Condition 1, 3 for Condition\leocw{}{s} 2 and 5, 2 for Condition 3, and 5 for Condition 4. The overall acceptance rate \leocw{}{across participants} was 60.4\% ($N=108$).

\begin{figure}
    \centering
    \includegraphics[width=0.5\textwidth,keepaspectratio, trim={0 24.5cm 5cm 0},clip]{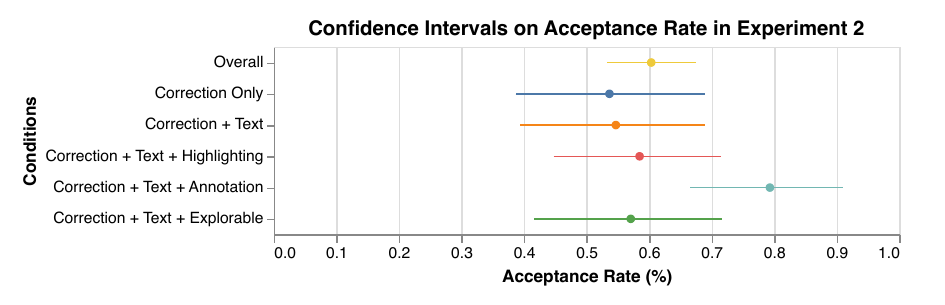}
    \vspace{-20pt}
    \caption{Confidence intervals of \leocw{}{the} acceptance rate in\leocw{ the}{} phase 2 of \leocw{experiment two}{Experiment Two}\leo{ suggesting the corrected version of the chart chosen by the participants in phase 1.}{, \ie, the percentage of accepting the suggested correct version of the chart after viewing the explanation.}}
    \label{fig:exp2ciplot}
    \squeezeafterfigure
\end{figure}

A post hoc pairwise t-test with Bonferroni correction revealed no significant differences across conditions. The confidence interval plot in \cref{fig:exp2ciplot} showed \leocw{}{a} high data variation. It remains to be seen whether adding explanations on top of presenting only the corrected version is more convincing or has no effect. Therefore, we cannot reject the null hypothesis of Hypothesis 2.1. However, the \leocw{annotation}{\annotation} condition exhibited a high mean acceptance rate, which could be a hint for further investigation.

Across different issues, \leocw{truncated axis}{Truncated Axis} had a significantly higher acceptance rate than the other two issues, vs. \leocw{rainbow colors}{Inappropriate Color Scheme} ($p: 0.007$) and vs. 3D pie charts ($p: 0.046$). Since we analyzed only three issues after excluding the two issues \leocw{with}{of} limited sample size, we cannot determine whether participants were more convinced by the \leocw{truncated axis explanations}{explanations on Truncated Axis} or were simply reluctant to accept the corrected \leocw{version}{versions}. Nonetheless, there was a difference in the acceptance rate across issue types, leading us to reject the null hypothesis of Hypothesis 2.2.

Since Hypothesis 2.1 is not supported, and Hypothesis 2.3 is a stronger claim than Hypothesis 2.1, Hypothesis 2.3 is also not supported. We cannot conclude that \leocw{explorable explanations}{\expexp} \leocw{have}{has} a stronger \leocw{convincing power in persuading}{persuasive power of convincing} users to change their design choices.

The experiment data and processing scripts can be found on OSF\footref{osf}.

\section{Discussion and Future Work}


Our data found little to no correlation between the SGL assessment score and the accuracy of identifying misleading charts, \leocw{either}{neither} in overall accuracy \leocw{or}{nor} accuracies \leocw{for}{of each} individual issue \leocw{types}{type} ($0 < r < 0.2$). The questions in the SGL assessment focus on the ability to interpret accurate charts\leocw{,}{} but do not assess the participants' ability to question the validity of the charts. In the OGL assessment \cite{galesic2011graph}, there are questions where the correct answer is "cannot infer from the \leocw{chart."}{chart," but this aspect of critical chart reading is neglected in SGL.}
\leo{}{Recently, Ge \etal have compiled a set of 45 assessment questions\leocw{,}{} designed to evaluate critical thinking abilities when confronted with misleading visualizations \cite{ge2023calvi}.}
We recommend that future studies on this topic consider modifying the SGL assessment by including one or two questions to address the weakness of SGL.


In \leocw{phase two}{the second phase} of \leocw{experiment two}{Experiment Two}, we collected participants' justifications for why they chose to keep their selection of a misleading chart over the accurate one. We coded the responses and analyzed the results. \leocw{For truncated axis ($N=58$), participants}{58 out of 186 responses rejected the correction on Truncated Axis. Participants} found that the lengths of the bars without truncating the axis were too similar and difficult to compare ($79.3\%, N=46$). Even though they were aware of the problem of exaggeration, they still preferred the chart with a truncated axis. Ritchie \etal also investigated the value of \leocw{}{the} truncated axis over \leocw{one showing}{the one that shows} the whole axis\leocw{ and}{. They} designed an interaction technique to allow readers to switch between the whole and \leocw{}{the} truncated axis \cite{ritchie2019lie}.

\leocw{For rainbow colors ($N=61$), people}{61 out of 138 responses rejected the correction on Inappropriate Color Scheme. Participants} reported that sequential colors were difficult to read, and rainbow colors offered better distinguishability and chart readability ($90.2\%, N=55$). One participant rejected the correction\leocw{ because}{:} ``It is much easier to differentiate between rainbow colours than different shades of one colour.'' \leocw{However, there}{There} were also responses stating that rainbow colors were helpful to colorblind people ($6.6\%, N=4$), even though rainbow colors are, in fact, very difficult to read for people with color weakness\leocw{}{es}. \leocw{For 3D pie charts ($N=63$), participants}{63 out of 175 responses rejected the correction on 3D pie charts. Participants} preferred 3D pie charts for aesthetic reasons ($41.3\%, N=26$) and believed that since the values were labeled, readers could always refer to the labels and not be misled by them ($34.9\%, N=22$). One participant wrote\leocw{}{,} ``Your explanation might be true\leocw{,}{.} However, \leocw{The}{the} number\leocw{}{s} are clearly marked.''

Some justifications are valid and worth considering when constructing explanations to address misunderstandings. The preference for high contrast (between colors or bar length) over accurate interpretation may be a design trade-off from the participants' perspective. Providing better alternatives and explaining the implications to users or readers can help them make more informed design decisions.
\leo{}{Although the guidelines for visualization design are generally considered best practice\leocw{}{s}, their application is subject to the context and the specific purpose of the visualization. If deviations from these guidelines can be justified logically, they can \leocw{indeed prove beneficial for}{benefit} the task in question. The active field of research aiming to understand human perception and the interpretation of visualizations is persistently questioning current practices, unveiling their shortcomings, and improving both visualization tools and guidelines.}


The main idea of this work is inspired by the \leocw{explorable explanations}{\expexp} developed for different topics across various fields\leocw{,}{} and \leocw{}{conjecturing that} applying this approach to explain visualization guidelines could be valuable. The work by Hopkins \etal implemented and evaluated the \leocw{highlighting}{\highlighting} explanation method \cite{hopkins2020visualint}. The experiment result is similar to our results in this study, showing no significant improvement when adding visual explanations on top of text explanations. The work by Fan \etal implemented the \leocw{correction}{\correction} explanation, and their evaluation supported the notion that people with access to explanations perform better \leocw{in}{at} identifying misleading visualizations \cite{fan2022annotating}. \leocw{However}{Though}, they did not compare it with the base case of text-only explanations. It remains an open question: \leocw{is}{Is} text explanation alone enough to explain misleading visualizations? We must admit that we did not find evidence to reject this idea. \leocw{However,}{Imagine that} if \leocw{}{the} classic books by Huff \cite{huff1993lie} and Tufte \cite{tufte1983visual} had used only text explanations, they might not have inspired many other authors to publish titles or blog posts on the topic\leocw{, and}{. Consequently,} the public might have \leocw{ended up}{become} even less aware of the pitfalls of misleading charts. 

One of the primary objectives of this study is to inform the future development of visualization recommendation systems and linting systems regarding the design of user interfaces when the system attempts to suggest a better alternative visualization to the user.
Although our experiment results did not conclude that \leo{annotation}{\leocw{explorable explanations}{\expexp}} has greater persuasive power over other explanation methods, it is worth noting that the design space of \leo{annotation}{\leocw{explorable explanations}{\expexp}} is much larger and highly dependent on the explanation authors' creativity. \leonotice{[Mistake in draft: should be explorable explanations instead of annotation.]}
Further studies may yield more conclusive results.

\section{Conclusion}

\leocw{In this study, we}{This study} identified six distinct explanation techniques through an extensive review of online resources, books, and a design workshop. We applied these methods to five chart-related issues, incorporating the concept of \leocw{explorable explanations}{\expexp}. Our two crowdsourced evaluation experiments revealed that exposure to explanations enhanced participants' ability to recognize misleading charts. However, no significant differences were observed among the various explanation techniques. \leocw{In the persuasiveness assessment, we discovered that participants were inclined to accept more than 60\% of the proposed adjustments.}{We discovered that participants were inclined to accept more than 60\% of the proposed adjustments in the persuasiveness assessment.}

Nevertheless, we found no significant differences among the explanation methods in convincing participants to accept the modifications.
\leo{Developing more effective explanation techniques to address the resistance to adopting less visually appealing sequential color schemes and 2D pie charts remains an ongoing challenge.}{It remains a challenge to develop more effective explanation techniques in educating and convincing the general audience on the application of visualization best practices.
At a minimum, they need to be made aware of the potential pitfalls and benefits when they decide not to follow these guidelines.
}


\bibliographystyle{abbrv-doi-hyperref}

\bibliography{template}

\end{document}